\documentclass[
    ,final            
  ]
  {aipproc}

\usepackage{indentfirst}

\layoutstyle{8x11single}

\begin{document}

\title{A Study of Cooling Time Reduction of\\Interferometric Cryogenic Gravitational Wave Detectors\\Using a High-Emissivity Coating}

\classification{04.80.Nn, 44.40.+a}

\keywords      {gravitational wave, emissivity, coating, cooling time, radiation }

\newcommand{\icrr}{Institute for Cosmic Ray Research (ICRR), University of Tokyo, \mbox{5-1-5 Kashiwanoha, Kashiwa, Chiba 277-8582, Japan}}
\newcommand{\kek}{High Energy Accelerator Research Organization (KEK), 1-1 Oho, Tsukuba, Ibaraki 305-0801, Japan}

\author{Y. Sakakibara}{
  address={\icrr}
}

\author{N. Kimura}{
  address={\kek}
}

\author{T. Suzuki}{
  address={\kek}
}

\author{K. Yamamoto}{
  address={\icrr}
}

\author{D. Chen}{
  address={\icrr}
}

\author{S. Koike}{
  address={\kek}
}

\author{\mbox{C. Tokoku}}{
  address={\icrr}
}

\author{T. Uchiyama}{
  address={\icrr}
}

\author{M. Ohashi}{
  address={\icrr}
}

\author{K. Kuroda}{
  address={\icrr}
}

\begin{abstract}
In interferometric cryogenic gravitational wave detectors,
there are plans to cool mirrors and their suspension systems (payloads) in order to reduce thermal noise, that is, one of the fundamental noise sources. 
Because of the large payload masses (several hundred kg in total) and their thermal isolation, a cooling time of several months is required.
Our calculation shows that a high-emissivity coating (e.g. a diamond-like carbon (DLC) coating) can reduce the cooling time effectively by enhancing radiation heat transfer.
Here, we have experimentally verified the effect of the DLC coating on the reduction of the cooling time.
\end{abstract}

\maketitle


\section{Introduction}
Gravitational waves, predicted by Einstein's theory of general relativity, have not been detected directly yet. The direct detection of such waves is of great importance in physics and astronomy. The first generation of interferometric gravitational wave detectors, such as LIGO\cite{LIGO}, VIRGO\cite{VIRGO}, GEO\cite{GEO}, and TAMA\cite{TAMA}, are already in operation, and several second-generation detectors are currently under construction, such as AdLIGO, AdVIRGO, and KAGRA (the nickname of the Large-scale Cryogenic Gravitational wave Telescope (LCGT), a Japanese km-scale cryogenic detector project)\cite{KAGRA}. It is expected that these second-generation detectors would be able to detect gravitational waves directly. Third-generation gravitational wave detectors, such as ET (the Einstein Telescope)\cite{ET}, are also being planned. KAGRA and ET enjoy two key advantages over other detectors: they will be located at an underground site with small seismic motion, and they are equipped with a cooled mirror (at around 20 K in the case of KAGRA) to reduce thermal fluctuation (thermal noise) of the mirror and its suspension\cite{uchi}.

In the interferometric gravitational wave detectors, the mirror is suspended because it has to be a free test mass and has to be isolated from seismic vibration.  In the cryogenic ones, such as KAGRA and ET, fibers will be used to extract heat from the mirror to keep the mirror cooled\cite{Basti}.  To reduce vibration via the fibers, the mass to be cooled should be large, and the fibers should be thin.  This is the main reason why it will take a long time to cool down cryogenic interferometric gravitational wave detectors.  Because the detectors cannot operate while they are cooling down, the long cooling time decreases the observation duty factor of the detectors.  Therefore, it is necessary to reduce the cooling time.

There are several methods for reducing the cooling time.
\begin{enumerate}
\item Increasing radiation: Radiation is enhanced by increasing the emissivity of the mass to be cooled.  Although this method is a less efficient means of heat transfer than the other two methods described next, it requires no apparatus and has a high stability appropriate for large-scale cooling.
\item Gas cooling: Heat is extracted from the mass to be cooled by gas exchange.  This method needs some device to open and close the holes in the radiation shield for the main laser beam and for the vibration isolation system.
Another problem with this method is that the evacuation speed of turbo-molecular pumps for helium gas, often used for this purpose, is slow.
\item Heat switch: When the detector is cooled down, a conductor of heat is attached to the mass to be cooled.  When the detector is operating, the conductor is removed.  This method also needs an apparatus to attach and detach to the payloads.
\end{enumerate}
The method of increasing the radiation has been chosen for KAGRA since it avoids the use of a mechanical apparatus, which could cause unstable operation of the cooling system.  This paper describes a reduction in cooling time by increasing radiation and its experimental verification for KAGRA.

\section{Calculation of cooling time}
\begin{figure}
\includegraphics[height=.3\textheight]{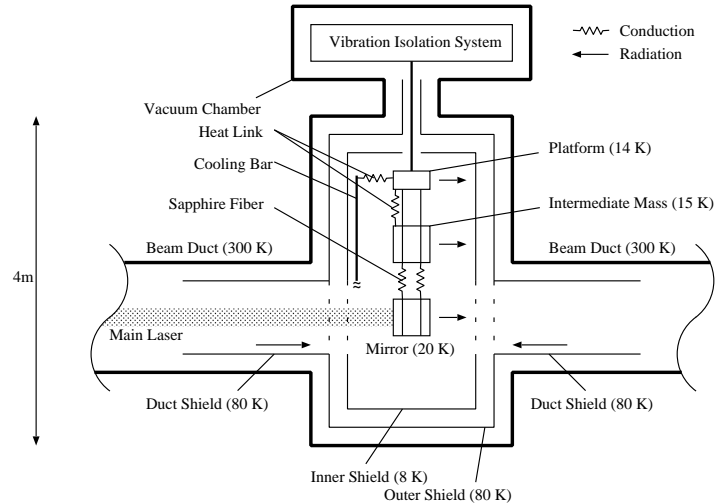}
\caption{Schematic diagram of one of KAGRA's cryostats. A sapphire mirror is suspended by thin sapphire fibers. Heat absorbed by the sapphire mirror is conducted to the intermediate mass by means of the sapphire fibers and then to the cooling bar through the platform by aluminum heat links.  There are four double-stage cryocoolers (not shown) for each cryostat used.  The cooling bar is connected to the second stages of the two cryocoolers (not shown). The inner shield is also connected to the second stages of the other two cryocoolers (not shown). The outer shield is connected to the first stages of the four cryocoolers.  There are two recoil masses (not shown) surrounding the mirror and the intermediate mass to control the positions of the two masses.  The duct shields will be installed to prevent thermal radiation from the holes for the laser beam\cite{duct}.
}
\label{lcgtshield}
\end{figure}

\begin{figure}
\includegraphics[height=.25\textheight]{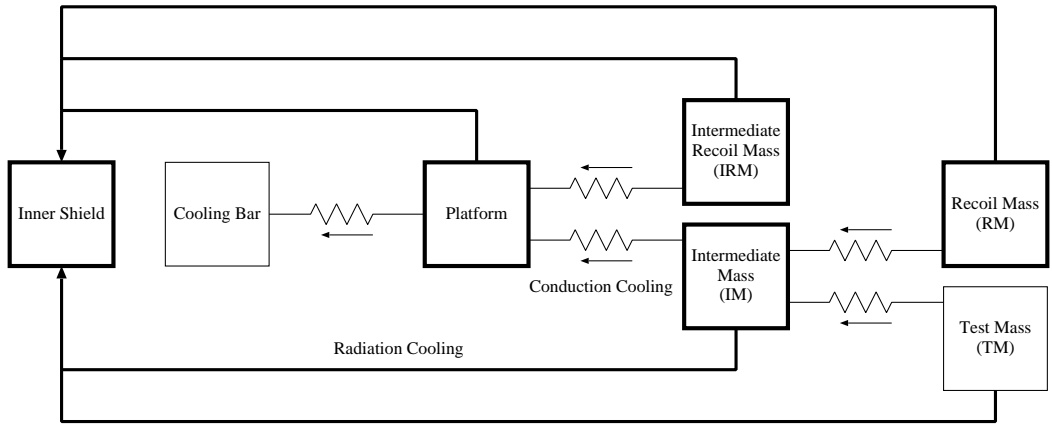}
\caption{Schematic diagram of a calculation model of the KAGRA cryogenic system.  Each payload is modeled as a point mass.  These point masses are thermally connected by thermal conduction and radiation.  The inner surface of the inner shield and all the payloads except the test mass (that is, the boxes with thick borders) will be coated with the high-emissivity coating.}
\label{calcmodel}
\end{figure}

\begin{table}
\begin{tabular}{cccc} \hline
& Material & Mass[kg]&Effective surface area[$\mathrm{m}^2$]\\ \hline
Test mass (TM)&Sapphire&22.8&0.076\\
Recoil mass (RM)&Copper&37.2&0.182\\
Intermediate mass (IM)&Copper&60.9&0.124\\
Intermediate recoil mass (IRM)&Copper&61.7&0.358\\
Platform&Copper&121&0.434\\ \hline
\end{tabular}
\caption{Design parameters of the KAGRA payloads.  Surfaces hidden by another payload are excluded from the effective surface area.}
\label{KAGRApayload}
\end{table}

\begin{table}
\begin{tabular}{cccccc} \hline
&& Material & Number & Diameter[mm] & Length[mm]\\ \hline
Cooling bar&Platform&Aluminum&7&1&785\\ 
Platform&IM&Aluminum&5&3&628\\ 
Platform&IRM&Aluminum&5&3&628\\ 
IM&TM&Sapphire&4&1.6&300\\ 
IM&RM&Aluminum&4&1.6&471\\ \hline
\end{tabular}
\caption{Design parameters of the fibers between the KAGRA payloads.}
\label{KAGRAfiber}
\end{table}

\paragraph{Method}
Figure \ref{lcgtshield} shows a schematic diagram of a cryostat used to cool one of the mirrors in KAGRA.  To reduce the seismic vibration, the mirror is suspended at the lowest part of the vibration isolation system, which is made of multi-stage pendulums.  The lowest five masses, called payloads (TM: test mass (mirror), RM: recoil mass, IM: intermediate mass, IRM: intermediate recoil mass, and platform), will be cooled down to reduce the thermal noise.

To calculate the cooling time, a model, where point masses are thermally connected by thermal conduction and radiation, was used and is shown schematically in Figure \ref{calcmodel}.  Parameters of each payload and the fibers between the payloads are given in Table \ref{KAGRApayload} and \ref{KAGRAfiber}.  The inner surface of the inner shield and all the payloads except the test mass will be coated with the high-emissivity coating (e.g. a diamond-like carbon (DLC) coating\footnote{Interferometric gravitational wave detectors require high vacuum because a fluctuation of the refractive index along the path of the main laser beam causes noise.  The coating used in the cryostats must be vacuum-compatible.  From this point of view, a DLC coating is a promising candidate\cite{vacuum}.}).  In this calculation, the temperature of the inner shield was taken from the measured value of the actual cryostat during the cooling test described later.  The temperature of the cooling bar was fixed at 8 K in this calculation.  The equation solved here is
\begin{equation}
\frac{dT}{dt}=\frac{Q(T)}{M C(T)}
\end{equation}
where $T$ is temperature, $t$ is time, $M$ is mass, and $C(T)$ is the specific heat at temperature $T$.  $Q$ is the sum of the heat radiated (described by eq.(\ref{eqrad})) and the heat transferred by conduction (described by eq.(\ref{eqcond})).  The heat radiated by the mass $x$ to the inner shield is (e.g. \cite{holman})
\begin{equation}
Q_x(T_x,T_\mathrm{sh})=\epsilon _xA_x\sigma (T_x^4-T_\mathrm{sh}^4), \label{eqrad}
\end{equation}
where $\sigma$ is the Stefan-Boltzmann constant, $\epsilon _x$ is the emissivity of the mass $x$, and $A_x$ is the effective surface area of mass $x$.  The effective surface area represents the surface area which contributes to thermal radiation.  Namely, surfaces hidden by another payload are excluded.  The surface area of the inner shield, described later, is sufficiently large compared to those of the payloads that the heat transfer is independent of the emissivity of the inner shield.  The radiation heat transfer between the payloads was neglected because the  temperature differences between the payloads are smaller than those between the payloads and the inner shield.  On the other hand, the heat conducted by $N$ fibers with temperatures $T_x$, $T_y$ at the ends, a cross-section of $S$, a length of $\ell $, and a thermal conductivity of $\kappa (T)$ is
\begin{equation}
Q_{xy}(T_x,T_y)=\frac{NS}{\ell}\int ^{T_y}_{T_x}\kappa (T')dT'.
\label{eqcond}
\end{equation}
Since the thermal conductivity of sapphire fibers is proportional to their diameter\cite{tomarufiber}, the value in ref.\cite{tomarufiber} multiplied by the ratio of the diameter of the fibers was used.  The other material values such as the thermal conductivity of the aluminum fibers and the specific heat of copper or sapphire were taken from literature (e.g, \cite{material}).  The emissivity of the high-emissivity coating (DLC was used here) and that of sapphire are
\begin{eqnarray}
\epsilon _\mathrm{DLC}&=&0.3\times \left( \frac{T}{300\ \mathrm{K}} \right) \label{emdlc} \\
\epsilon _\mathrm{sap}&=&0.5 \label{emsap},
\end{eqnarray}
which were obtained by our experiments, as described later.  The emissivity for copper without any coating was assumed to be 0.03.

\begin{figure}
\includegraphics[height=.3\textheight]{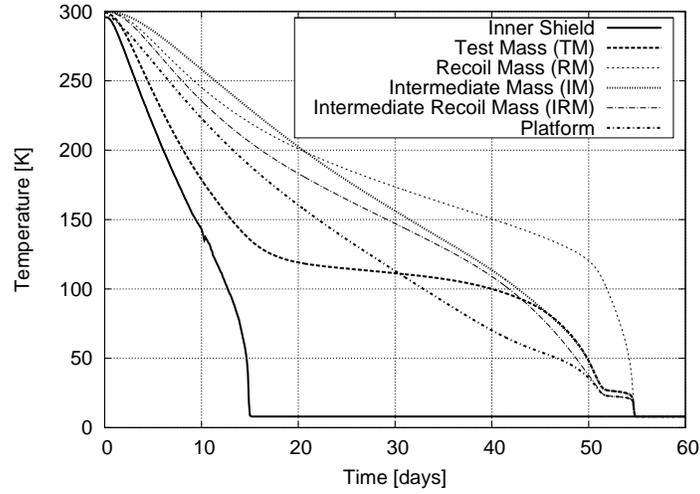}
\caption{Calculation results of cooling time of the KAGRA payloads without any coating.
}
\label{cooling11nct}
\end{figure}

\begin{figure}
\includegraphics[height=.3\textheight]{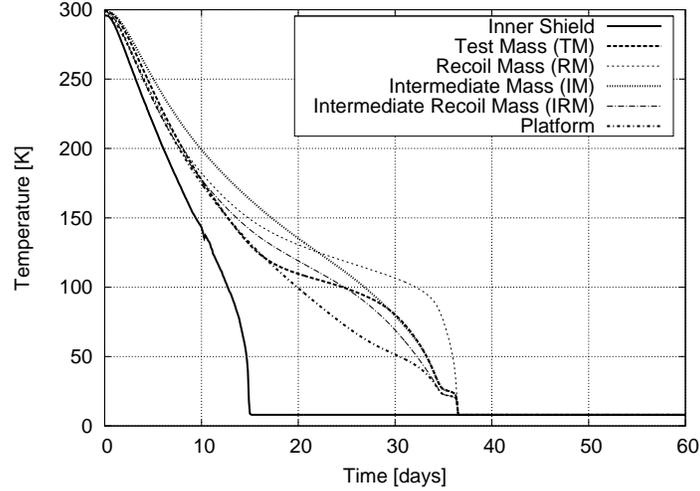}
\caption{Calculation results of the cooling time of the KAGRA payloads with the DLC coating.}
\label{cooling11ct}
\end{figure}

\paragraph{Results}
The calculation results are shown in Figures \ref{cooling11nct} and \ref{cooling11ct}.  Above approximately 150 K, the payloads are cooled mainly by thermal radiation.  Below roughly 150 K, the conduction cooling is dominant.  The DLC coating can reduce the cooling time when the radiation cooling is dominant.  While it takes two months to cool down completely without any coating, the DLC coating can significantly reduce this time to only one month.

\section{Experiments}
In order to verify the effect of the high-emissivity coating on the reduction of the cooling time, experiments were conducted with half-sized payloads and two spheres.

\begin{figure}
\includegraphics[height=.25\textheight]{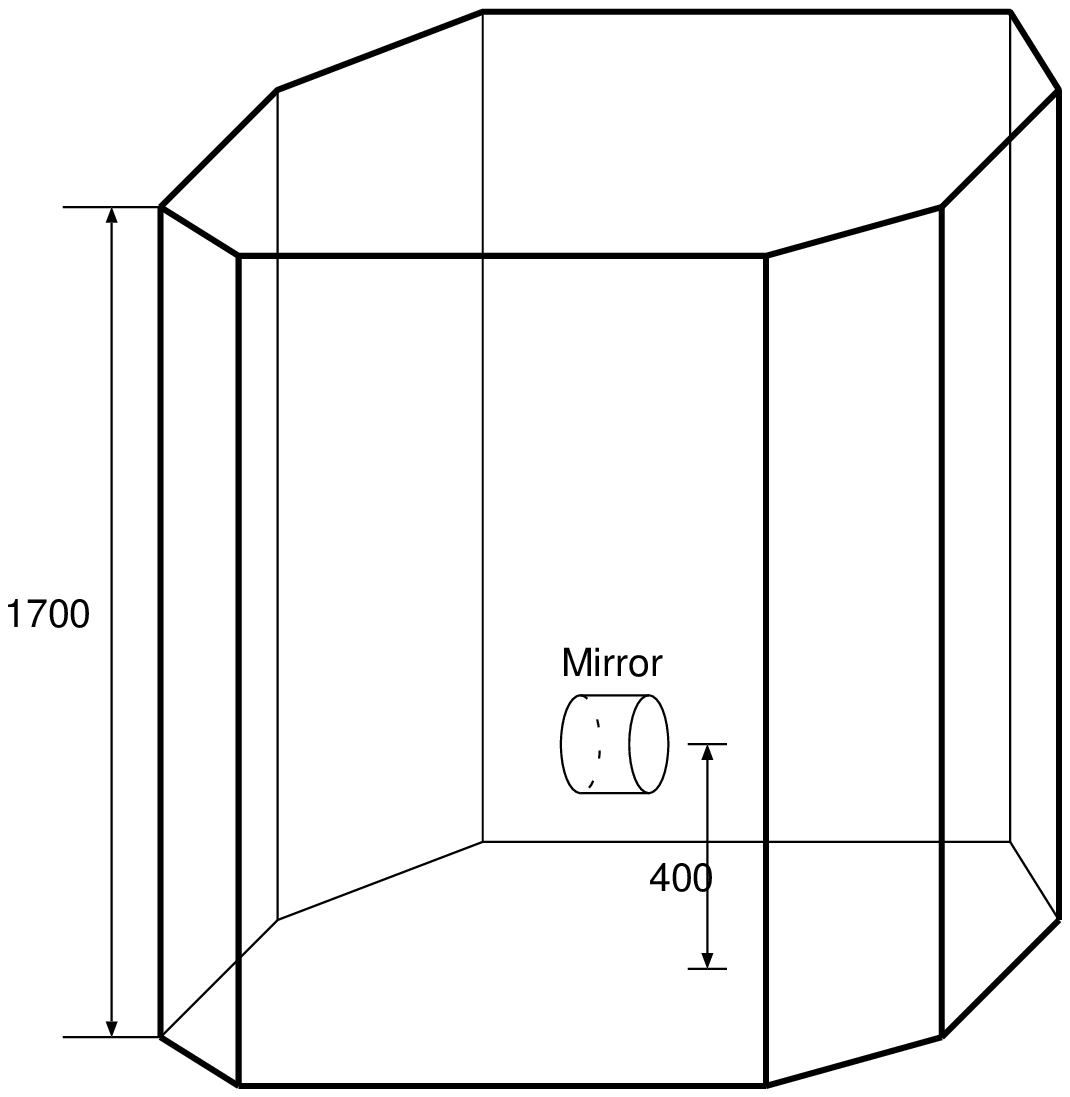}
\ \ \ \ \ \ \ \
\includegraphics[height=.2\textheight]{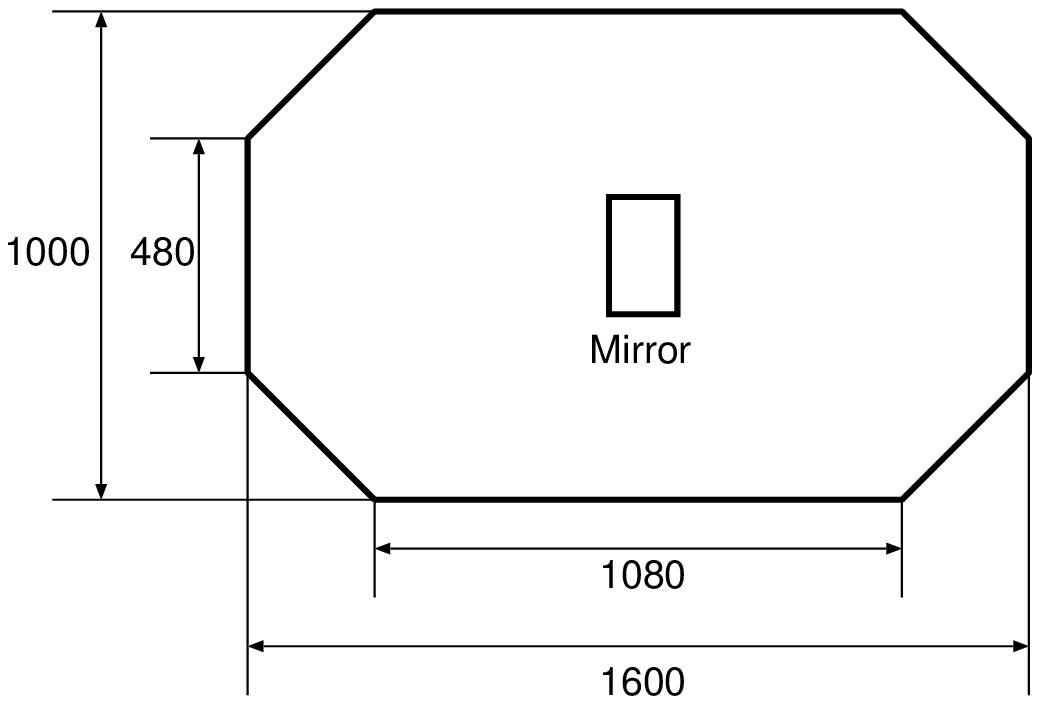}
\caption{
Schematic diagram of the inner shield of the KAGRA cryostat.  The picture on the right-hand side is a top view of the inner shield (units: mm).
}
\label{kagrashield}
\end{figure}

\begin{figure}
\includegraphics[height=.3\textheight]{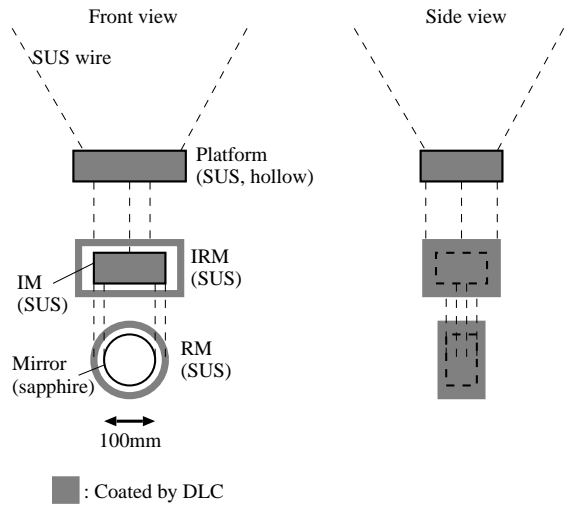}
\caption{Schematic diagram of the half-sized payloads.  All the payloads except the mirror are coated with a DLC.  The payloads were suspended inside the inner shield by SUS wires.
}
\label{diagram_dummy}
\end{figure}

\paragraph{Method}
The KAGRA cryostat has double radiation shields, an inner shield and an outer shield, made of aluminum A1070.  Figure \ref{kagrashield} shows an inner shield.  The inner shield, the size of which is limited by the maximum transportation size on Japanese public roads, is designed to be as large as possible to allow humans to work inside the shield.  The DLC coating covered approximately 50\% of the surface area of the inner surface of the inner shield.
The main purpose of this DLC coating is to absorb scattered light from the mirror.
 
A cooling test of the KAGRA cryostats was conducted\cite{tokoku}.  During the cooling test, half-sized payloads, shown in Figure \ref{diagram_dummy}, were suspended in the inner shield of one of the cryostats.  Spheres with and without the DLC coating were suspended inside the inner shields of other two cryostats to focus on examining the emissivity of aluminum with and without the DLC coating.  The holes for the laser beam and for the vibration isolation system were closed by aluminum plates to block out the 300 K thermal radiation.  After the pressure reached the order of $10^{-3}\ \mathrm{Pa}$, the cryocoolers were turned on, and the cooling of the inner shield began.

The half-sized payloads in the first cryostat were designed to have half the size of the actual KAGRA payloads to reduce the cooling time (their parameters are shown in Table \ref{dummypayload}).  All the payloads except the mirror are made of SUS304 and coated with DLC.  Thermometers (DT-670-CU made by LakeShore, not calibrated) were attached to each payload.  The payloads were suspended within the inner shield by SUS wires, the thermal conduction of which is negligible.  The mirror of the half-sized payloads was at the position where the actual KAGRA mirror would be suspended (400 mm in height).  Neither the aluminum heat links nor the sapphire fibers were used since this experiments concentrated on an analysis of the radiation cooling.

The spheres, suspended in the second and the third cryostat, are made of aluminum A5000 lathed and chemically polished (CP) with a diameter of 105 mm.  One has no coating, and the other one is coated with DLC on the CP surface.  Each sphere is hollow and made of two hemispheres, where the material inside the hemispheres has been removed.  The total mass of each sphere is 1.2 kg.  Thermometers (DT-670-CU made by LakeShore, not calibrated) were attached to each hemisphere.  The sphere was suspended at the position where the actual KAGRA mirror would be suspended (400 mm in height) within the inner shield.  As suspension wires, four Kevlar wires (commercial aramid fibers), the thermal conduction of which is negligible, were used.

\paragraph{Results}

\begin{figure}
\includegraphics[height=.3\textheight]{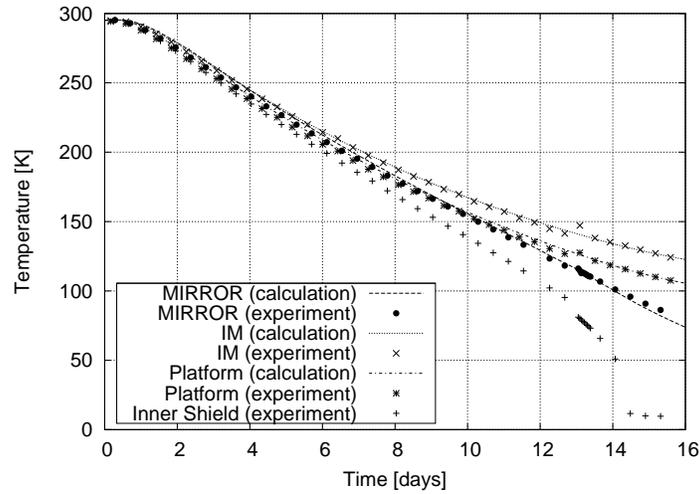}
\caption{Results of the experiments in the KAGRA cryostats using the half-sized payloads.  The calculated temperature of the payloads is the value calculated from the temperature of the inner shield.
}
\label{toshibaresult3}
\end{figure}

\begin{table}
\begin{tabular}{cccc} \hline
&Material &Mass[kg]&Effective surface area[$\mathrm{m}^2$]\\ \hline
Test mass (TM)&Sapphire&1.87&0.0157 \\
Recoil mass (RM)&SUS&3.72&0.0440 \\
Intermediate mass (IM)&SUS&5.18&0.0434\\
Intermediate recoil mass (IRM)&SUS&4.54&0.0743\\
Platform&SUS&4.75&0.109\\ \hline
\end{tabular}
\caption{Parameters of the half-sized payloads. }
\label{dummypayload}
\end{table}

\begin{figure}
\includegraphics[height=.3\textheight]{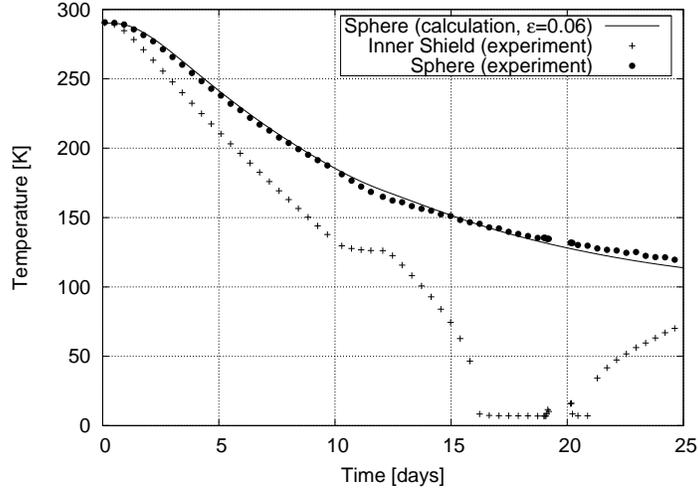}
\caption{Results of the experiments in the KAGRA cryostats using the sphere without any coating.  The temperature of the inner shield stayed constant between the 10th and the 12th day because the power supply was turned off.  On the 21st day, the cryocoolers were turned off in order to heat up the cryostat.  The calculated temperature of the sphere is the value calculated from the temperature of the inner shield, including the constant temperature between the 10th and the 12th day.
}
\label{toshibaresult1}
\end{figure}

\begin{figure}
\includegraphics[height=.3\textheight]{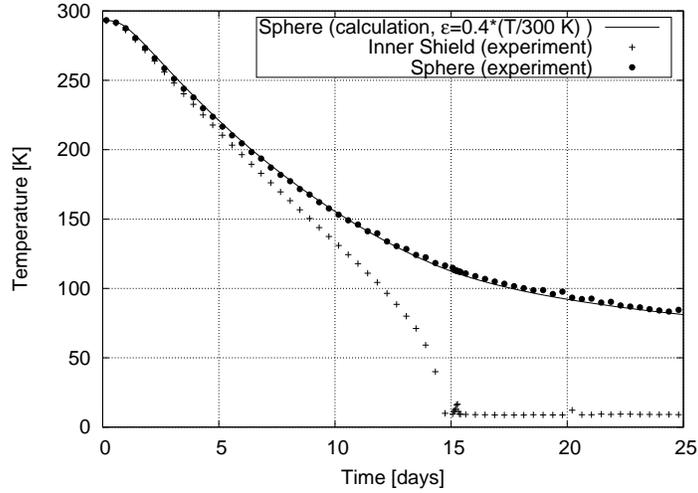}
\caption{Results of the experiments in the KAGRA cryostats using the sphere with the DLC coating.  The calculated temperature of the sphere is the value calculated from the temperature of the inner shield.
}
\label{toshibaresult2}
\end{figure}

The results of the experiments are shown in Figures \ref{toshibaresult3}-\ref{toshibaresult2}.  The temperature of the payloads and the spheres was also calculated from the temperature of the inner shield using eq.(\ref{eqrad}).  The sphere with the DLC coating cools faster than the sphere without any coating.  The measured results fit well to those of the calculation.

\section{Discussion}
The values for the emissivity of the sphere without any coating and that of the sphere with the DLC coating used to fit the results were, 
\begin{equation}
\epsilon _\mathrm{sphere,no\ DLC}=0.06, \quad
\epsilon _\mathrm{sphere,DLC}=0.4\times \left( \frac{T}{300\ \mathrm{K}} \right) \label{emdlc2}.
\end{equation}
In the temperature region we are interested, namely, above 150 K, where thermal radiation is dominant in cooling of the KAGRA payloads, the sphere with the DLC coating has a higher emissivity than the sphere without any coating.  Here, we have verified the effect of the DLC coating on the reduction of the cooling time.  The emissivity of the sphere with a DLC coating is proportional to the temperature $T$.  This fact can be explained as follows: emissivity is equal to the absorptivity at the wavelength of radiation $\lambda$.  The wavelength is inversely proportional to $T$ ($\lambda=10\ \mu\mathrm{m}$ at $T=300\ \mathrm{K}$).  When the coating has a constant absorption coefficient, the absorptivity is proportional to the coating thickness $d$ (equal to approximately 1 $\mu$m in these experiments) and inversely proportional to $\lambda$.  Namely, the coating looks thinner for longer wavelengths.  Thus, the absorptivity is proportional to $T$.  Similarly, the emissivity values of the payloads were obtained to be
\begin{equation}
\epsilon _\mathrm{Platform,DLC}=0.3\times \left( \frac{T}{300\ \mathrm{K}} \right) , \quad
\epsilon _\mathrm{IM,DLC}=0.4\times \left( \frac{T}{300\ \mathrm{K}} \right), \quad 
\epsilon _\mathrm{TM,sap}=0.5 \label{emdlc3}.
\end{equation}
Eqs. (\ref{emdlc2}), (\ref{emdlc3}) give several different values for the emissivity of the DLC coating.  This can be caused by differences in the thickness of the coating or the surface treatment of the metal.  Among these values, the smallest value $0.3\times ( T/300\ \mathrm{K})$ was used in the calculation of the cooling time of the KAGRA payloads as shown in eq.(\ref{emdlc}).  The emissivity of sapphire obtained here was used in eq.(\ref{emsap}).

In the current plan of KAGRA, the shields are cooled down only by the cryocoolers.  Cooling of the payloads with the DLC coating is limited by cooling of the inner shield above approximately 150 K, as shown in Figure \ref{cooling11ct}.  Although cryogens, such as liquid nitrogen, can help the cooling of the shields, there is a higher risk for gas from the cryogens to fill the underground environment.  It is also useful to look for material which has both higher emissivity than the DLC coating and vacuum compatibility to use in cryogenic interferometric gravitational wave detectors.

\section{Conclusion}
We calculated the cooling time of the KAGRA payloads using the emissivity of the DLC coating obtained from the experiments.  The calculation results show that the DLC coating can reduce the cooling time of the KAGRA payloads from two months to one month.  The experiments have verified the calculation model and the effect of a high-emissivity coating on the reduction of the cooling time.
Since KAGRA cannot operate during the cooling down, a long cooling time decreases the observation duty factor of KAGRA.  Even if the DLC coating can reduce the cooling time to approximately one month, cooling down once per year still occupies roughly 10\% of the available time.  Research on and development of another method is necessary to make the cooling time even shorter in the future.


\begin{theacknowledgments}
We thank Toshiba Corporation for manufacturing the KAGRA cryostats and helping us with these experiments.
Y. Kobayashi made the half-sized payload and the spheres used in this experiment.
A. Khalaidovski gave us useful comments on this manuscript.
This work was supported by the Leading-edge Research Infrastructure Program from the Ministry of Education, Culture, Sports, Science and Technology (MEXT).
\end{theacknowledgments}



\bibliographystyle{aipproc}   

\bibliography{sample}

\IfFileExists{\jobname.bbl}{}
 {\typeout{}
  \typeout{******************************************}
  \typeout{** Please run "bibtex \jobname" to optain}
  \typeout{** the bibliography and then re-run LaTeX}
  \typeout{** twice to fix the references!}
  \typeout{******************************************}
  \typeout{}
 }


\end{document}